\documentclass{emulateapj}

\catcode`\"=\active\let"=\"

\def\3{{\ss} }

\def\c12{{1\over 2}}

\def\plusplus{\raise 0.3ex\hbox{${\scriptstyle ++}$}{}}

       \def\down#1{\leavevmode \lower.70ex\hbox{#1}}
        
        \def\gta{\mathrel{\down{$\buildrel > \over \sim$}}} 

\newcommand{\oversim}[2]{\protect{\mbox{\lower0.5ex\vbox{%
   \baselineskip=0pt\lineskip=0.2ex
   \ialign{$\mathsurround=0pt #1\hfil##\hfil$\crcr#2\crcr\sim\crcr}}}}} 
\begin{document}

\title{On the formation of extended galactic disks by tidally disrupted dwarf galaxies}
\author{Jorge Pe\~{n}arrubia
, Alan McConnachie \& Arif Babul }
\affil{University of Victoria, 3800 Finnerty Rd., Victoria, BC, V8P 5C2, Canada
}
\begin{abstract}
We explore the possibility that extended disks, such as that recently
discovered in M31, are the result of a single dwarf
($10^9$--$10^{10}M_\odot$) satellite merger. We conduct N-body
simulations of dwarf NFW halos with embedded spheriodal stellar
components on co-planar, prograde orbits in a M31-like host galaxy.
As the orbit decays due to dynamical friction and the system is
disrupted, the stellar particles relax to form an extended,
exponential disk-like structure that spans the radial range
$30$--$200$ kpc. The disk scale-length $R_d$ correlates with the
initial extent of the stellar component within the satellite halo: the
more embedded the stars, the smaller the resulting disk
scale-length. If the progenitors start on circular orbits, the
kinematics of the stars that make up the extended disk have an average
rotational motion that is $ 30 - 50$\,km/s lower than the host's
circular velocity.  For dwarf galaxies moving on highly eccentric
orbits ($e \simeq 0.7$), the stellar debris exhibits a much lower
rotational velocity. Our results imply that extended galactic disks
might be a generic feature of the hierarchical formation of spiral
galaxies such as M31 and the Milky Way.
\end{abstract}


\section{Introduction}\label{sec:int}

Within the context of the hierarchical formation scenario, galaxies
such as M31 are expected to have assimilated 100--500 smaller mass
systems over a Hubble time (e.g.~Moore et al~1999). Today, stars from
these systems are expected to be found in the disk, bulge and stellar
halo (e.g Abadi et al. 2003, 2006, Governato et al. 2004 and
references therein). The fossil signatures of these remnants are a
veritable treasure trove of information about the galaxy assembly
process and the nature of the merging subunits.

Detailed investigations of M31 (Ibata et al. 2001, 2004, 2005,
Ferguson et al.  2002, 2005, McConnachie et al.  2003, 2004, Zucker et
al.  2004, Fardal et al. 2006, Font et al.  2006, Guhathakurta et
al. 2006, Kalirai et al. 2006, Brown et al. 2006, Chapman et al. 2006)
reveal a large number of fossil substructures.  Intriguingly,
many of these have disk kinematics and are distributed in a gigantic
flattened structure surrounding the high-surface brightness inner
disk (Ibata et al.~2005).  The observations are consistent with this
extended structure being a rotating disk-like system whose
midplane is coincident with that of the inner disk, but which
extends $\gta 80$ kpc from the center of M31.  The rotational
velocities of the constituent stars lag the M31 disk rotation by $\sim
30$\,km/s (Ibata et al.~2005).  The mass of the extended disk is $\sim
10\%$ of the inner disk and its stellar surface density profile of the
structure is consistent with a exponential decline although there is
some uncertainty about the value of the radial scale length (cf.
Ibata et al~2005; Kalirai et al.~2006).  Here, we explore whether the
extended disk could be the remnant of a dwarf galaxy.

\section{Constructing our galaxy models}

\subsection{The host galaxy}\label{sec:host}

Our M31 galaxy model consists of three static sub-components: a
Miyamoto-Nagai (1975) disk, a spherical Hernquist (1990) bulge and a
spherical Navarro, Frenk \& White (1997) (NFW) dark matter halo. The
corresponding potentials are:

\begin{eqnarray}
\Phi_d  & = & -\frac{G M_d}{\sqrt{R^2+(a+\sqrt{z^2+b^2})^2}} \label{eq:phid};\;\;\;\;
\Phi_b  =  -\frac{G M_b}{r+c};\; \nonumber\\
\label{eq:phih} 
\Phi_h & = & -\frac{G M_h \ln(1+r/r_s)}{r}\big[\ln(1+c_{200})-\frac{c_{200}}{1+c_{200}}\big]^{-1}.
\end{eqnarray}

\noindent Following Geehan et al.~(2005), we set $M_d=8.4\times
10^{10}M_\odot$, $M_b=3.3 \times 10^{10} M_\odot$, $a=5.4$ kpc,
$b=0.26$ kpc and $c=0.61$ kpc. The M31 halo parameters are
$M_h(r_{200})=6.8\times 10^{11}M_\odot$, $r_{200}=180$ kpc, $r_s=8.18$
kpc, which yields a concentration of $c_{200}=r_{200}/r_s=22$.

\subsection{Generating N-body satellite galaxies in equilibrium}\label{sec:generate}

Our satellite models have two components: (i) a spherical NFW dark
matter halo, (ii) an embedded spherical ``stellar'' component that
reproduces a King (1966) density distribution and approximates the
light profile of the dSph galaxies of the Milky Way (MW) and M31
(Irwin \& Hatzidimitriou 1995, McConnachie \& Irwin 2006).

We use the code of Kazantzidis et al.~(2004, 2006) to generate NFW
satellite halos with $10^6$ particles, $c=r_{\rm vir}/r_s=20$ and an
exponential cut-off at the virial radius.  We calculate the relative
energy, $\epsilon=\Psi-v^2/2$, and the relative potential,
$\Psi=-\Phi_{NFW} + \Phi_{NFW}(\infty)$, for all satellite
particles. Next, we generate $n$ bins in energy and, for each bin, we
select $f_\star(\epsilon)/f_{NFW}(\epsilon)\times N(\epsilon)$
``stellar'' particles, where $N(\epsilon)$ is the number of NFW
particles within $(\epsilon,\epsilon+d\epsilon)$ and
$d\epsilon=\epsilon_{\rm max}/n$. Here, $f_\star$ and $f_{NFW}$ are the
isotropic distribution function for the stellar and the dark matter
particles calculated from

\begin{eqnarray}
f_i(\epsilon)=\frac{1}{8\pi^2}\bigg[\int_0^\epsilon \frac{d^2\rho_i}{d\Psi^2}\frac{d\psi}{\sqrt{\epsilon-\Psi}}+\frac{1}{\sqrt{\epsilon}}\bigg(\frac{d\rho_i}{d\Psi}\bigg)_{\Psi=0}\bigg],
\label{eq:edding}
\end{eqnarray}

\noindent where the subsript $i$ denotes that $\rho$ can be either a NFW or a
King profile.  The result is $N_\star$ dark matter particles with a
density distribution

\begin{eqnarray}
\rho_\star=\frac{K}{x^2}\bigg[\frac{\cos^{-1}(x)}{x}-\sqrt{1-x^2}\bigg],~~x\equiv\frac{1+(r/r_c)^2}{1+(r/r_t)^2}~.
\label{eq:rhok}
\end{eqnarray}

\noindent $r_c$ and $r_t$ are the core and tidal radii, $K$ is an arbitrary
constant, and we have fixed $r_t/r_c=5$ (see Bullock \&
Johnston~(2005) for more details). By construction,
the stellar component does not contribute to the potential of the
satellite galaxy, nor does it influence its evolution. 

\subsection{Satellite structural and orbital parameters}\label{sec:dwarf_mod}

We consider satellites with mass $M[r_{\rm vir}]=5\times 10^{9}
M_\odot$ (model H1), the estimated mass of typical Local Group dwarf
galaxies, and $M[r_{\rm vir}]=5\times 10^{10} M_\odot$ (model H2), at
the top end of minor mergers in merger trees for a M31-like
galaxy. Different King profiles are used for the tracer stellar
particles, characterized by the quantity $r_c/r_s$. This determines
the compactness of the stellar component in the halos. We select
$r_c/r_s=1.0, 0.5, 0.1$ (K1, K2 and K3, respectively).  The number of
stellar particles available to trace the King profiles decreases as
$r_c/r_s$ decreases, so that $N_\star=3.1\times 10^5, 1.3\times 10^5,
1.5\times 10^4$ for K1, K2 and K3, respectively.

\begin{figure}
\plotone{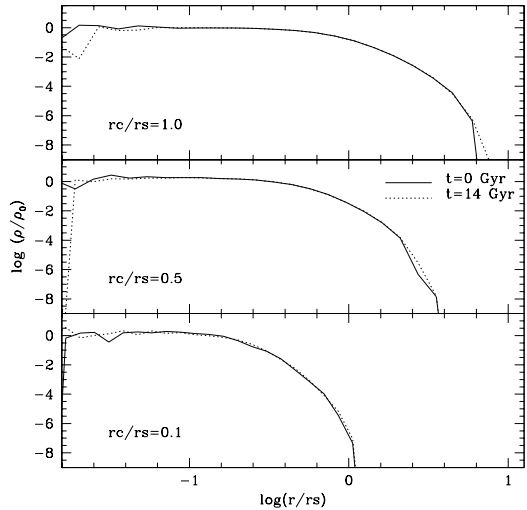}
\caption{ Density profile of the stellar component normalized to the
central censity ($\rho_0$). Solid and dotted lines show respectively
the profiles at $t=0$ and after 14 Gyr of evolution in isolation (i.e
in absence of external forces). Our method to embed a ``stellar''
profile within a NFW halo produces systems in equilibrium.}
\label{fig:dens_is}
\end{figure}

Satellites are placed in the host galaxy at an initial apocenter of
$r_a=75$ kpc. The velocity is chosen to obtain the desired orbital
eccentricity. We explore (i) an initially
circular orbit, and (ii) a highly eccentric orbit
($e=[r_a-r_p]/[r_a+r_p]=0.71$, or $r_a/r_p=1/6$, where $r_a$ and $r_p$
are the apo and peri center distances). The latter corresponds to the
typical eccentricity of satellite galaxies in cosmological simulations
(Ghigna et al. 1998).  For this Letter, we only study satellites whose
orbits are co-planar and prograde with the M31 disk.

\section{N-body simulations}

We refer the reader to Pe\~narrubia et al.~(2006) for a detailed
description of our N-body code and the numerical parameters and
techniques used in our simulations. Fig.~\ref{fig:dens_is} shows the
density profiles of K1, K2, K3 at $t=0$ (solid lines) and after
evolving them for 14 Gyr in isolation. This shows that (i) the method
outlined in \S\ref{sec:generate} produces N-body systems in perfect
equilibrium (to within Poisson fluctuations) and (ii) the selection of
the N-body code's numerical parameters is appropriate.

\section{Results}\label{sec:results}

\subsection{Disruption of satellite galaxies}\label{sec:xy}

\begin{figure}
\plotone{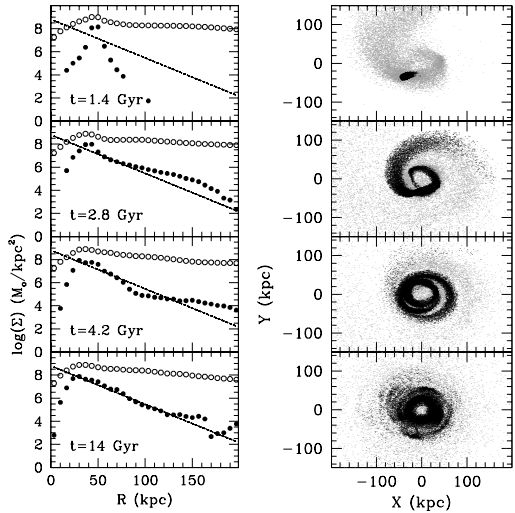}
\caption{ {\it Left column}: Surface density as a function of radius
for the model K2H2 on an initially circular orbit. Stellar and dark
particles are shown with filled and open circles, respectively. Each
panel corresponds to a different time. The solid lines shows the
surface density of an exponential disk $\Sigma=\Sigma_0 \exp(-R/R_d)$
with $R_d=30$ kpc. {\it Right column}: Distribution of stellar (black
dots) and dark matter (grey dots) debris at the same snap-shots. Note
how remarkably the stellar debris resemble an exponential disk after
the satellite galaxy has been totally disrupted.}
\label{fig:xyt}
\end{figure}

The process of disruption occurs as the satellite galaxy sinks to the
central regions of the host galaxy (owing to dynamical friction) and
tidal forces strip the dark and stellar particles. The sequence for
model K2H2 can be seen in Fig.~\ref{fig:xyt}: the left column shows
the surface density profile of stars (filled circles) and dark matter
(open circles) at different times. The right column shows the
projection of stellar (black dots) and dark matter (grey dots)
particles onto the disk plane of the host galaxy.  The top panel shows
the satellite galaxy at a time when most of the stellar material is
still bound (see the corresponding right-hand panel), so that the
surface density profile is fairly peaked at the satellite location. As
time passes, the stellar particles start to form tidal streams, which
can be seen as bumps in $\Sigma(R)$. As the debris configuration
relaxes (mixes in phase-space), the curve $\Sigma(R)$ smooths and
approaches an exponential-like surface density profile over the
interval $30\le R \le 200$\,kpc. The solid line in the left panels
shows the profile of an exponential disk with $\Sigma=\Sigma_0
\exp(-R/R_d)$ and $R_d = 30$\,kpc.  Since the stellar component
is initially tightly bound, then the stellar surface density
distribution is considerably steeper than that of dark matter
particles.

The time-scale for the extended disk to form is a combination of the
decay and relaxation times ($t_{\rm dec}$ and $t_{\rm rel}$,
respectively). For massive satellites (eg. Fig.~\ref{fig:xyt}), the
satellite sinks and disrupts within $t_{\rm dec}\simeq 3$\,Gyr. Thus,
the relaxation and decay times are only comparable in the inner
regions of the host galaxy. Debris at $R\gtrsim120$ kpc are not
relaxed even after a Hubble time.

\subsection{Spatial distribution of debris}\label{sec:dens}

\begin{figure}
\plotone{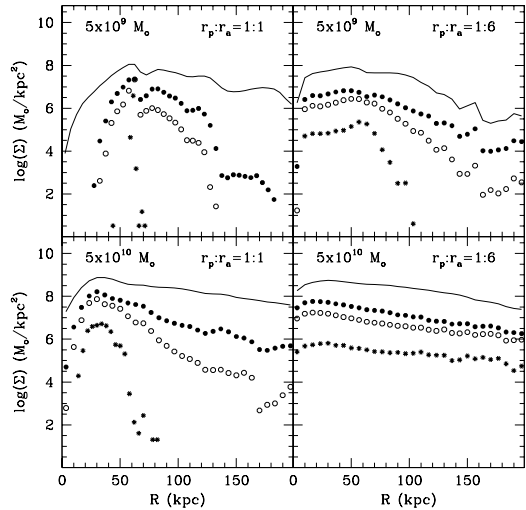}
\caption{ Surface density of debris as a function of radius after 14
Gyr of evolution. Models K1, K2 and K3 are represented by filled
circles, open circles and stars, respectively. Left and right columns
show the distribution of debris of a galaxy on a circular and a highly
eccentric orbit, respectively.  Strong solid lines show the
distribution of dark matter particles for comparison. Note that the
resulting stellar density depends on both, the dwarf orbit and its
initial distribution within the halo.}
\label{fig:dens}
\end{figure}

While the above results are for a specific case, the emergence of an
extended exponential disk appears to be a generic feature of the
disruption of satellites on co-planar orbits. In Fig.~\ref{fig:dens},
we show the surface density profiles of the stellar and dark matter
particles after 14\,Gyr of evolution for various combinations of
satellite masses, embedded stellar systems and orbital
eccentricities. The left and right columns show satellite galaxies on
circular and highly eccentric orbits, respectively. Top and bottom
panels are for satellite galaxies with initial masses of $5\times 10^9
M_\odot$ and $5\times 10^{10} M_\odot$, respectively.

In the case of a lower-mass satellite on an initially circular orbit,
the surface density of stellar debris is not smooth even after 14 Gyr
and does not resemble exponential. This is because the satellite has a
long orbital decay time. The system has disrupted relatively recently
and the debris has not yet relaxed. This is in contrast to the case of
a massive satellite galaxy on a highly eccentric orbit (left-bottom
panel). The dynamical friction is so efficient that the satellite
sinks to the galaxy center and disrupts within $\sim 2$ Gyr. Since
there is no difference in the disruption time-scales for the dark and
stellar particles, both distributions are practically the same: an
exponential with a large radial scale $R_d\simeq 105$ kpc, independent
of how compact the stellar system was initially.

The intermediate cases in the bottom-left (massive satellite, circular
orbit) and the top-right (average-mass dwarf, eccentric orbit) panels
show exponential stellar profiles over finite radius: $30 -
120$\,kpc for the circular orbit, and $60 - 150$\,kpc for the
eccentric one. We find that the more compact the stellar component,
the steeper the final stellar surface density profile. The
scale-lengths of the exponential profiles are $R_d \simeq 50, 30$ and
7\,kpc for models with initial $r_c/r_s=1.0, 0.5 $ and 0.1,
respectively, independent of initial orbital eccentricity.

\subsection{Debris kinematics}\label{sec:kin}

\begin{figure}
\plotone{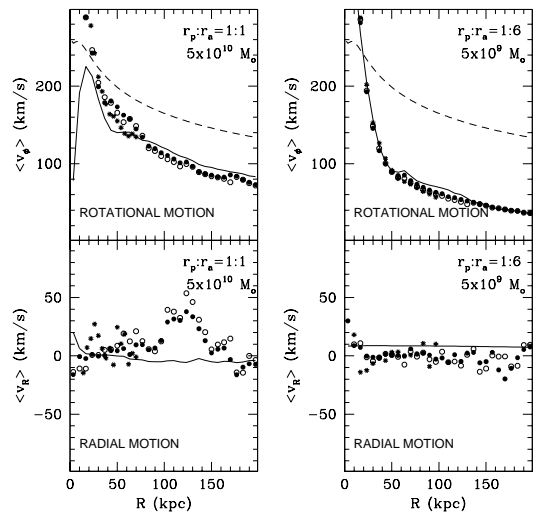}
\caption{ Average azimuthal (top panels) and radial (bottom panels)
stellar velocities as a function of radius for models K1, K2, K3 (same
notation as in Fig.~\ref{fig:dens}). The left and right columns show
satellites moving on a circular and a highly eccentric orbit,
respectively.  Note that the velocities of stellar debris are
insensitive to their initial distribution within the dwarf halo but
they reflect the orbital properties of the progenitor galaxy.}
\label{fig:vel}
\end{figure}

The extended disk in M31 was primarily identified by its rotating
kinematic signature. In Fig.~\ref{fig:vel} we plot the average radial
and azimuthal velocities of the satellite debris for those simulations
that form an exponential-like disk after 14\,Gyr (left
column: massive satellite on a circular orbit; right column:
typical-mass satellite on an eccentric orbit). For comparison, we also
show the circular velocity curve of the host galaxy (dashed lines).

From the panels showing the mean galactocentric radial velocities, we
see that the remnant of both satellites are practically relaxed, in
that the stream motion in the radial direction is small. More
interestingly, the extended debris exhibits a net rotation about the
host galaxy center, independent of the satellite's initial orbital
eccentricity.  For the satellite galaxy on an initially circular
orbit, the rotational velocity of the debris is $\sim 20 - 50$\,km/s
lower than the host circular velocity for $R \gta 30$ kpc. The
satellite on an initially eccentric orbit shows a much lower
rotational velocity, typically about 100\,km/s lower than $v_c$ for $R
\gta 40$ kpc.  Also, the stellar kinematics trace the dark matter
debris, independent of the initial distribution of stars within the
dwarf galaxy.

\section{Implications}

We find that stellar debris, resulting from the disruption of dwarf
galaxies on prograde orbits which are co-planar with the host galaxy
disk, relax into an extended exponential disk.  The disk scale-length
is strongly correlated with the initial distribution of stars inside
the dwarf halo, such that more embedded stellar profiles have smaller
``disk'' scale-length. For the simulations shown here, $R_d = 50, 30$
and $7$\,kpc for the initial stellar compactness parameter $r_c/r_s =
1.0, 0.5$ and $0.1$, respectively.  The stellar debris exhibit
rotation, even in the case of dwarf galaxies on initially highly
eccentric orbits. The average rotational velocity is around $30 -
50$\,km/s lower than the host circular velocity if the satellite
initially moves on a circular orbit. For highly eccentric orbits, the
resulting rotational velocity is a factor $\sim 2$ lower.

In the specific case of M31, Ibata et al.~(2005) find the extended
disk of M31 has a rotational velocity that is $\sim 30$\,km/s lower
than $v_c$. Based on our results, the observed kinematics favour a
scenario where the disk progenitor was a massive satellite galaxy on a
low-eccentricity, low-inclination orbit, possibly similar to the
progenitor of the Monoceros stream in the MW (Pe\~narrubia et
al. 2005).

How common should such extended disks be?  $\Lambda$CDM simulations
find that spiral galaxies experience several mergers of massive, $0.01
- 0.1\,M_{\rm host}$ dwarf galaxies (e.g. Gao et al. 2004), many on
low eccentricity orbits (Ghigna et al. 1998). That extended disks
appear aligned with the inner disk is not surprising taking into
account the strong decrease in orbital inclination suffered by massive
satellites in flattened systems (Quinn \& Goodman 1986, Pe\~narrubia,
Kroupa, Boily 2002), as well as orbital circularization (Jiang \&
Binney 2000).  Thus, in a $\Lambda$CDM work frame, our results
indicate that extended disks might be a common result of the
hierarchical formation of spiral galaxies such as the MW and M31.

Would accretion destroy the inner disk?  Our simulations assume a
static, non-responsive host galaxy potential. Several works (Toth \&
Ostriker 1992, Walker, Mihos \& Hernquist 1996, Velazquez \& White
1999) have shown that, for sufficiently massive satellites, the inner
disk would be disrupted by the merger event. To analize this
possibility we have carried out self-consistent N-body simulations of
the infalling systems considered in this {\em Letter}. Our preliminar
results indicate that the inner disk is not strongly affected by these
accretion events since, owing to the nearly exponential mass-loss rate of
NFW halos, the dwarf galaxies are disrupted {\it before} dynamical
frictions brings these systems close to the inner disk. The results
from these simulations, together with the analysis of the response of
the inner disk, will be shown in a subsequent paper.

\vskip0.5cm We thank Stelios Kazantzidis for the use of his code,
and Scott Chapman, Julio Navarro, Raja Guhathakurta and Mark Fardal
for useful discussions.  JP thanks Julio Navarro for financial support. AB was supported by NSERC through the
Discovery Grant program. AM thanks Julio Navarro and Sara Ellison for
financial support.

{}

\end{document}